\documentstyle[prl,aps,epsf]{revtex}
\begin{document}
\advance\textheight by 0.2in
\draft
\twocolumn[\hsize\textwidth\columnwidth\hsize\csname@twocolumnfalse
\endcsname  

\title{Positional Disorder (Random Gaussian Phase Shifts) in the
Fully Frustrated Josephson Junction Array (2D XY Model)}

\author{Pramod Gupta\cite{R0} and S. Teitel}

\address{Department of Physics and Astronomy, University of Rochester, 
Rochester, NY 14627}

\date{\today}	

\maketitle

\begin{abstract}
We consider the effect of positional disorder on a Josephson junction array
with an applied magnetic field of $f=1/2$ flux quantum per unit cell.
This is equivalent to the problem of random
Gaussian phase shifts in the fully frustrated 2D XY model.
Using simple analytical arguments and numerical simulations, we 
present evidence that the ground state vortex lattice of the
pure model becomes disordered, in the thermodynamic limit, by {\it any} 
finite amount of positional disorder.
\end{abstract}


]

The stability of vortex lattices to random disorder
is a topic of considerable recent interest, motivated by  
studies of the high temperature superconductors.  
In two dimensions (2D),
periodic arrays of Josephson junctions
form a well controlled system for investigating
similar issues of vortex fluctuations and disorder.  Here
we consider the effect of ``positional'' disorder 
on the vortex lattice of the fully 
frustrated Josephson array, with $f=1/2$ flux quantum of applied 
magnetic field per unit cell.  

Positional disorder \cite{R1,R2,R6,R20} was first discussed
with respect to the Kosterlitz-Thouless (KT) transition for 
the $f=0$ model in zero 
magnetic field.  Early arguments \cite{R1} predicting a reentrant normal phase 
at low temperatures have been revised by recent works \cite{R2,R20} which 
argue that there is a finite critical disorder strength 
$\sigma_c\simeq\sqrt{\pi/8}$; 
for $\sigma<\sigma_c$ an ordered 
state persists for $0\le T\le T_c(\sigma)$.  
For the pure $f=1/2$ case on a square grid \cite{R7,R3}, 
the ordered state has two broken symmetries: 
the $U(1)$ symmetry (``KT-like'' order) associated 
with superconducting phase coherence, and the $Z(2)$ symmetry 
(``Ising-like'' order) associated with the ``checkerboard'' vortex 
lattice, in which a vortex sits on every other site.  Previous
works \cite{R4,R5} have considered the effect of positional disorder on this $f=1/2$
model; all have concluded that both Ising-like and
KT-like order persist for at least small disorder strengths
$\sigma$.  In this work, however, we present new
arguments that suggest that, 
for $f=1/2$, the critical disorder is $\sigma_c=0$.

The Hamiltonian for the Josephson array is given by the 
``frustrated'' 2D XY model \cite{R3}, 
\begin{equation}
    {\cal H}[\theta_i] = \sum_{i\mu}U(\theta_i-\theta_{i+\hat\mu}-A_{i\mu})
	\label{eq:HXY}\enspace ,
\end{equation}
where $i$ are the sites of a periodic square grid with basis
vectors $\hat\mu=\hat x$, $\hat y$,
the sum is over all nearest neighbor (n.n.) bonds $\langle i,i+\hat\mu\rangle$
and $\theta_i-\theta_{i+\hat\mu}-A_{i\mu}$
is the gauge invariant phase difference across 
the bond, with $A_{i\mu}=(\phi_0/2\pi)\int_i^{i+\hat\mu}{\bf A}
\cdot d\ell$ the integral of the vector potential.

Positional disorder arises from random geometric distortions of the bonds of 
the grid, resulting in, $A_{i\mu}=A_{i\mu}^{(0)}
+\delta A_{i\mu}$; $A_{i\mu}^{(0)}$ is the value in the absence
of disorder, and $\delta A_{i\mu}$ is the random deviation.  
We take the $\delta A_{i\mu}$
to be independent Gaussian random variables with
\begin{equation}
	[\delta A_{i\mu}]=0,\quad{\rm and}\quad [\delta A_{i\mu}\delta
    A_{j\nu}]=\sigma^2\delta_{ij}\delta{\mu\nu}
	\label{eq:dAij}\enspace.
\end{equation}
$[\ldots]$ denotes an average over the quenched disorder.
The positionally disordered array is thus also referred to as the
XY model with random Gaussian phase shifts.

When $U(\phi)$ is the Villain function \cite{R7.5}, 
the Hamiltonian ($\ref{eq:HXY}$) is equivalent to a dual ``Coulomb
gas'' of interacting vortices \cite{R7,R8,R18},
\begin{equation}
	{\cal H}[n_i]={1\over 2}\sum_{ij}(n_i-f-\delta f_i)G_{ij}
    (n_j-f-\delta f_j)\enspace .
	\label{eq:HCG}
\end{equation}
The sum is over all pairs of {\it dual} sites $i$, $j$, 
$n_i$ is the integer vorticity on site $i$, and the interaction $G_{ij}$ is
the Green's function for the 2D discrete Laplacian operator,
$\Delta_{ik}G_{kj}=-2\pi\delta_{ij}$, where $\Delta_{ij}\equiv
\delta_{i,j+\hat x}+\delta_{i,j-\hat x}+\delta_{i,j+\hat y}+
\delta_{i,j-\hat y}-4\delta_{ij}$.  For large separations, $G_{ij}
\simeq - \ln |{\bf r}_i-{\bf r}_j|$.  The
$f_i\equiv f+\delta f_i$ are $(1/2\pi)$ times the circulation of 
the $A_{i\mu}$ around dual site $i$; $f$ is the average
applied flux, while $\delta f_i$ is the
deviation due to the random $\delta A_{i\mu}$,
\begin{equation}
  	\delta f_i = {1\over 2\pi}\left[\delta A_{i,x}+\delta A_{i+\hat x,y}-
      \delta A_{i+\hat y,x}-\delta A_{i,y}\right]\enspace.
  	\label{eq:dfi}
\end{equation}
Geometrically distorting a bond increases the flux through the 
cell on one side of the bond, while reducing the flux through
the cell on the opposite side by the same amount.  
The $\delta f_i$ are thus {\it anticorrelated} among 
n.n. sites. Positional disorder is thus the same as
random dipole pairs of quenched charges $\pm \delta f_i$ \cite{R1}.  
From Eqs.~(\ref{eq:dAij}) and (\ref{eq:dfi}) we get,
\begin{equation}
	[\delta f_i]=0,\quad {\rm and}\quad [\delta f_i \delta 
    f_j]=-{\sigma^2\over 4\pi^2}\Delta_{ij}\enspace .
	\label{eq:df}
\end{equation}

The Hamiltonian (\ref{eq:HCG}) can be rewritten as interacting
charges in a one body random potential \cite{R2},
\begin{equation}
	{\cal H}[q_i]={1\over 2}\sum_{ij}q_iG_{ij}q_j-\sum_i q_iV_i\enspace,
	\label{eq:Hq}
\end{equation}
where $q_i\equiv n_i-f$, and the random potential is $V_i=\sum_j 
G_{ij}\delta f_j$.  For $f=1/2$, $q_i=\pm 1/2$.
From Eq.~(\ref{eq:df}),
\begin{eqnarray}
	[V_i]=0, &\quad& {\rm and}\quad [V_iV_j]=\sum_{k,l}G_{ik}[\delta f_k
    \delta f_l]G_{lj}    \nonumber\\
     &=&-{\sigma^2\over 4\pi^2}\sum_{k,l}G_{ik}\Delta_{kl}G_{lj}
    ={\sigma^2\over 2\pi} G_{ij}\enspace,
	\label{eq:dV}
\end{eqnarray}
The $V_i$ thus have logarithmic long range correlations.

We now use an Imry-Ma \cite{R9} type argument to estimate the stability of
the doubly degenerate checkerboard ground state to the formation of
a square domain of side $L$.  The energy of such an 
excitation consists of a domain wall term, $E_d$, which is present for the
pure case, and a pinning term, $E_p$, due to the interaction
with the random $V_i$.  $E_d(L)$ has the form \cite{R10},
\begin{equation}
	E_d\simeq aL+c\ln L+d\enspace.
	\label{eq:Ed}
\end{equation}
The first term is the interfacial tension of the domain wall;
the second term comes from net charge that builds up
at the corners of the domain \cite{R11}.  Calculating $E_d(L)$
numerically for a pure system, we find an excellent fit to 
Eq.~(\ref{eq:Ed}), with $a=0.28$, $c=0.15$, and $d=0.058$.

By Eq.~(\ref{eq:dV}), 
the average pinning energy of the domain ${\cal D}$, $[E_p]=
2\sum_{i\in {\cal D}}q_i[V_i]=0$, but the variance is,
\begin{equation}
	[E_p^2]=4\sum_{i,j\in {\cal D}}q_i[V_iV_j]q_j={4\sigma^2\over 2\pi}
    \sum_{i,j\in {\cal D}}q_iG_{ij}q_j={4\sigma^2\over \pi}E_0\enspace,
	\label{eq:Ep}
\end{equation}
where $E_0=(\pi/32)L^2$ is the ground state energy of the 
checkerboard domain \cite{R12}.  The root mean square pinning
energy is thus,
\begin{equation}
	[E_p]_{rms}=b L,\quad b={\sigma\over 2\sqrt 2}\simeq 
    0.35\sigma\enspace.
	\label{eq:Eprms}
\end{equation}
For domains whose energy is {\it lowered} by the interaction 
with $V_i$, the typical excitation energy is 
$E=E_d-[E_p]_{rms}$.   Eqs.~(\ref{eq:Ed}) and 
(\ref{eq:Eprms}) imply that when $b>a$, i.e. when
$\sigma>\sigma_c\simeq 0.8$, $E(L)$ has a maximum  at 
$L=\xi\equiv (c/2\sqrt 2)/(\sigma -\sigma_c)$.  Domains of size
$L>\xi$ will lower their energy by increasing in size, and so
disorder the system.  Thus, one naively expects that 
when $\sigma<\sigma_c$ the system
preserves its Ising-like order, but when $\sigma>\sigma_c$ the
system is disordered into domains of typical size $\xi$.

However the leading size dependencies of Eqs.~(\ref{eq:Ed}) and 
(\ref{eq:Eprms}), $E_d\sim [E_p]_{rms}\sim L$, are exactly the
same as found in the 2D n.n. random field Ising model (RFIM).  
For the RFIM it is known \cite{R9,R16,R13} that 2D
is the lower critical dimension, that the randomness causes
domains walls at $T=0$ always to roughen and so acquire an
effective {\it negative} line tension, and that the
critical disorder is $\sigma_c=0$, i.e. {\it any amount of
disorder, no matter how weak, destroys the Ising-like order of the
pure case}.  By analogy, we suggest that the positionally
disordered $f=1/2$ 2D XY model similarly has $\sigma_c=0$.  
Our conclusion, that $[E_p]_{rms}\sim L$ as in the 2D RFIM,
follows from a subtle cancellation between the long range interactions
between charges $q_i$, and the long range correlations of the
random potential $V_i$.

To check this prediction, we carry out 
Monte Carlo (MC) simulations of the Hamiltonian (\ref{eq:HCG})
with periodic boundary conditions on $L\times L$ square grids.  
Our MC procedure is as follows \cite{R12}.  One MC excitation attempt
consists of the insertion of a neutral $n=\pm 1$ vortex pair on n.n.
or next n.n. sites, which is accepted
or rejected using the usual Metropolis algorithm.  $L^2$
such attempts we call one MC pass.  At each temperature we 
typically used $4000$ MC passes to equilibrate the system,
followed by $128,000$ MC passes to compute averages.
Every $100$ passes we attempt a global excitation 
reversing the sign of all the charges, $q_i\to -q_i$.
For each disorder realization we cooled
down two distinct ``replicas'', starting with different 
random charge configurations and using different random number 
sequences.  In only about
$3\% $ of the cases did the two replicas fail to give reasonable
agreement.

To test for Ising-like order we define an order parameter
analogously to an Ising antiferromagnet,
\begin{equation}
	M={1\over L^2}\sum_i q_i(-1)^{x_i+y_i}\enspace.
	\label{eq:M}
\end{equation}
\begin{figure}
\epsfxsize=3.2truein
\epsfbox{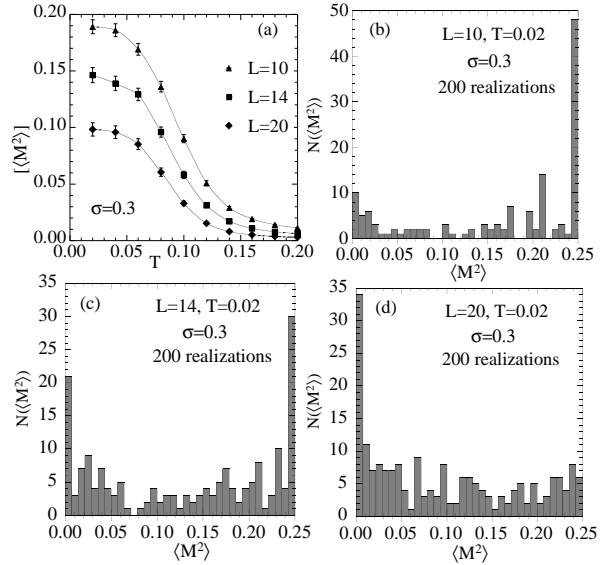}
\vspace{9pt}
\caption{(a) $[\langle M^2\rangle]$ vs. $T$, for $\sigma=0.3$,
and sizes $L=10$, $14$, $20$ (solid lines are guides to
the eye); Histogram of occurrences of
$\langle M^2\rangle$ in $200$ realizations of
disorder, for $\sigma=0.3$ at low $T=0.02$, for (b) $L=10$,
(c) $L=14$ and (d) $L=20$.
}
\label{f1}
\end{figure}
We first consider $\sigma=0.3$, smaller than both
the naive estimate of $\sigma_c=0.8$, and 
the $\sigma_c=\sqrt{\pi/8}\simeq 0.63$ of the 
$f=0$ model.   Fig.~\ref{f1} plots $[\langle 
M^2\rangle]$ vs. $T$, averaged over $200$ disorder realizations,
for sizes $L=10$, $14$ and $20$.  All curves
start to increase from zero near $T\simeq 0.13$, 
which is $T_c(\sigma=0)$ of the pure model.
However $[\langle M^2\rangle]$ at low $T$
decreases steadily with increasing $L$.  The reason for this
becomes clearer if we consider the histogram of values of $\langle 
M^2\rangle$ that occur as we sample the different realizations of disorder.
We show such histograms in Figs.~\ref{f1}b-d, for the lowest 
temperature $T=0.02$.  As $L$ increases, the statistical
weight shifts from predominantly ordered systems ($M^2=1/4$), to 
predominantly disordered systems ($M^2=0$).  Assuming that this
trend continues, we expect that as $L\to\infty$,  $[\langle 
M^2\rangle]\to 0$.  

\begin{figure}
\epsfxsize=3.2truein
\epsfbox{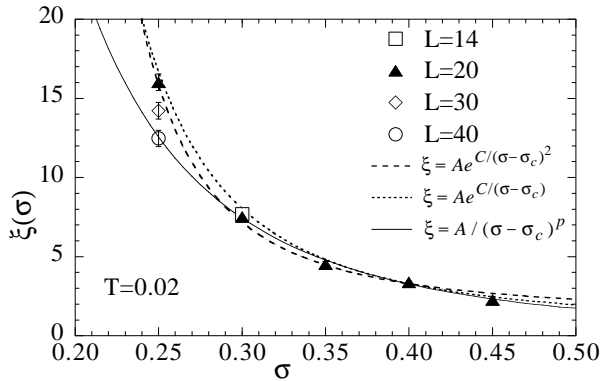}
\vspace{9pt}
\caption{
Correlation length $\xi(\sigma)$ vs. $\sigma$ at $T=0.02$ for various
sizes $L$.  Sizes $L=14, 20$ are averaged over $200$ 
realizations of the randomness; sizes $L=30, 40$ are averaged
over $50$ realizations.  Dashed, dotted and solid lines
are fits to the scaling forms (i), (ii) and (iii)
respectively.
}
\label{f2}
\end{figure}

To measure the ``random field
correlation length'' $\xi$, we consider the vortex correlation
function,
\begin{equation}
	S({\bf k})={1\over L^2}\sum_{i,j}e^{i{\bf k}\cdot ({\bf r}_i
    -{\bf r}_j)}\langle n_i n_j\rangle\enspace.
	\label{eq:S}
\end{equation}
For the pure case, $S({\bf k})$ in the ordered phase has
singular Bragg peaks at ${\bf K}=\pm\pi\hat x\pm\pi\hat y$.
If the vortex lattice is disordered, these peaks will broaden,
and their finite width provides a measure of $\xi$.  Writing
${\bf k}={\bf K}+\delta{\bf k}$, and assuming a Lorentzian
shape for the disorder averaged peak, $[S({\bf k})]
\propto 1/(\delta k^2 +\xi^{-2})$,
we determine $\xi$ by fitting to this form for $\delta k=0$, and
$\delta k=2\pi/L$ \cite{R15}.  In Fig.~\ref{f2}
we show $\xi$ vs. $\sigma$ at our lowest $T=0.02$, for several 
system sizes $L$.  Only for our smallest value $\sigma=0.25$ 
does a finite size effect remain.  In this case, however,
$\xi$ {\it decreases} as $L$ increases.  This is in contrast
to the increase of $\xi$ with $L$ that one would expect if one were
approaching a second order transition.  This behavior 
is consistent with that seen in
Figs.~\ref{f1}b-d, where as $L$ increases, a greater fraction
of the disorder realizations result in disordered states.  

We next fit our results for $\xi(\sigma)$
to several possible scaling expressions: (i) $\xi\sim 
e^{C/(\sigma-\sigma_c)^2}$, (ii) $\xi\sim e^{C/(\sigma-\sigma_c)}$,
and (iii) $\xi\sim |\sigma-\sigma_c|^{-p}$.
The first has been suggested by Binder \cite{R16} for the 2D RFIM.
While in Binder's expression $\sigma_c=0$, here we leave it as an
arbitrary parameter to be determined from the fit.  
The second has been suggested for the positionally disordered
$f=0$ model \cite{R1,R2}, in which $\sigma_c>0$.  
The third is the familiar power law form. 
Using data for only the largest $L$ for each 
$\sigma$, the results of these fits are shown in Fig.~\ref{f2}.
The value of $\sigma_c$ and the $\chi^2$ of the fit for each case is
(i) $\sigma_c=0.0046\pm 0.050$, $\chi^2=67$;(ii) $\sigma_c=0.0134\pm 
0.055$, $\chi^2=67$;
(iii) $\sigma_c=0.0013\pm 0.098$, $p=2.86\pm 0.84$, $\chi^2=7.6$. 
The power law (iii) gives a significantly better fit
than (i) or (ii), however all give $\sigma_c=0$ within
the estimated error.
Given the rather limited range of the data,
the above fits should be treated with caution.  However they
do indicate that the data contains no suggestion of a diverging
$\xi$ at a finite $\sigma$.
Coupled with our Imry-Ma argument, we thus find a consistent
picture suggesting that $\sigma_c=0$ for the $f=1/2$ 2D XY model.

\begin{figure}
\epsfxsize=3.2truein
\epsfbox{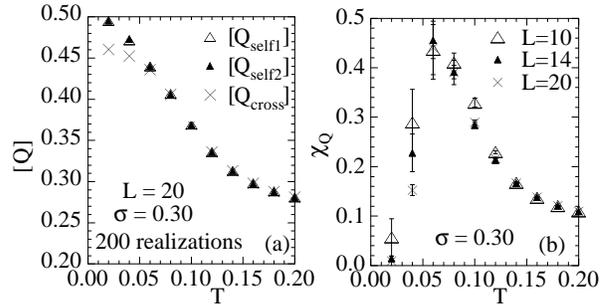}
\vspace{9pt}
\caption{(a) Overlaps $[Q_{\rm self\, 1}]$, $[Q_{\rm self\, 2}]$,
and $[Q_{\rm cross}]$ vs. $T$ for $\sigma=0.3$ and $L=20$;
(b) Overlap susceptibility $\chi_Q$ vs. $T$ for $\sigma=0.3$ 
and $L=10$, $14$, $20$.  Both are averaged over $200$ disorder realizations.
}
\label{f3}
\end{figure}

Returning to the case $\sigma=0.3$, where
Ising-like order has been lost,
we now consider whether the system may still 
have a finite temperature ``spin glass'' transition to 
a disordered but frozen vortex state.  To test for this
we measure the self and cross overlaps \cite{R17}, $Q_{\rm self}$ and $Q_{\rm 
cross}$, 
\begin{eqnarray}
    Q_{\rm self\, \alpha}&=&{1\over L^2}\sum_i\langle n_i^{(\alpha)}(t)
     n_i^{(\alpha)}(t+\tau)\rangle\nonumber\\
    Q_{\rm cross}&=&{1\over L^2}\sum_i\langle n_i^{(\alpha)}(t)
     n_i^{(\beta)}(t)\rangle\enspace.	
	\label{eq:Q}
\end{eqnarray}
$\alpha$ and $\beta$ index the two independent replicas.
For $\tau$ sufficiently large
we expect $Q_{\rm self\, 1}=Q_{\rm self\, 2}=Q_{\rm cross}$, 
if the system is well equilibrated. Averaging Eq.~(\ref{eq:Q})
over several values of $\tau\ge 2000$ to improve our statistics,
we plot $[Q_{\rm self\, 1}]$, $[Q_{\rm self\, 2}]$ and $[Q_{\rm cross}]$
vs. $T$ in Fig.~\ref{f3}a.  We see that our
system is fairly well equilibrated down to the lowest $T$ we study.
To test for a spin glass transition, we measure the overlap 
susceptibility,
\begin{equation}
	\chi_Q=L^2\left\{[\langle Q_{\rm cross}^2\rangle]-
     [\langle Q_{\rm cross}\rangle^2]\right\}
    \enspace,
	\label{eq:chiQ}
\end{equation}
which we plot vs. $T$ in Fig.~\ref{f3}b for various system sizes.
The peak in $\chi_Q$ near $T\simeq 0.06$ shows
no noticeable increase as $L$ increases, thus suggesting that
there is no finite temperature spin glass transition.

\begin{figure}
\epsfxsize=3.2truein
\epsfbox{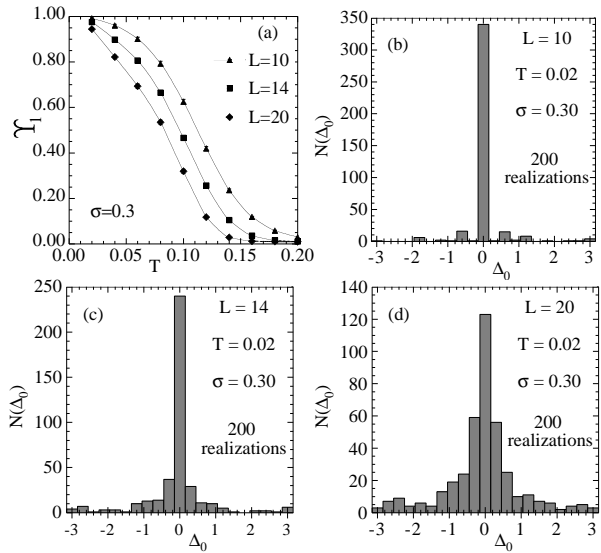}
\vspace{9pt}
\caption{(a) $\Upsilon_1$ vs. $T$, for $\sigma=0.3$,
and sizes $L=10$, $14$, $20$ (solid lines are guides
to the eye); Histogram of
values of $\Delta_0$ found in $200$ disorder realizations,
for $\sigma=0.3$ at low $T=0.02$, at (b) $L=10$,
(c) $L=14$, and (d) $L=20$.
}
\label{f4}
\end{figure}

If the vortices are not frozen, but are free to diffuse,
one expects that superconducting
phase coherence is also destroyed.  To explicitly test this
we measure the helicity modulus.  
The Hamiltonian (\ref{eq:HCG}) can viewed
as representing the XY model with ``fluctuating twist'' boundary
conditions \cite{R18}.  Using the method of Ref.~\cite{R19}, we  
determine the dependence of the total free energy $F$ of the corresponding
XY model, as a function of the twist $(\Delta_x,\Delta_y)$ 
which is applied in a ``fixed twist'' boundary condition.
We then determine the $(\Delta_{x0},\Delta_{y0})$ that
minimizes $F$; the helicity modulus tensor is then the curvature
of $F$ at the minimizing twist, $\Upsilon_{\mu\nu}=\partial^2F/
\partial\Delta_\mu\partial\Delta_\nu$.  In Fig.~\ref{f4}a
we plot $\Upsilon_1$, the largest of the two eigenvalues of $\Upsilon_{\mu\nu}$,
vs. $T$, for $\sigma=0.3$ and sizes $L=10,14,20$.  
At all $T$, $\Upsilon_1$ continues to decrease as $L$ increases,
giving no suggestion of a finite temperature 
transition.  In Figs.~\ref{f4}b-d we plot histograms of the minimizing
twist $\Delta_0$ for the three sizes $L$.  Note, in choosing our
random phase shifts $\delta A_{i\mu}$, we impose the constraint
$\sum_i\delta A_{i\mu}=0$ in order to remove one trivial source
of $\Delta_0\ne 0$.  We see that the width of the distributions
of $\Delta_0$ steadily increases with increasing $L$, suggesting 
\cite{R20}
that the strength of the random disorder is renormalizing to greater values
on larger length scales. 

To conclude, our results suggest that Ising-like order is destroyed
for any finite amount of positional disorder.  Further, we found
in one specific case that when the Ising-like order vanished,
no spin glass order or phase coherence existed either.  We speculate
that this remains true as well for any finite disorder strength.
Although $\sigma_c=0$, the finite $\xi(\sigma)$ nevertheless can
become extremely large for small values of $\sigma$.  When $\xi$
exceeds the size of the experimental or numerical sample, the system
will indeed look ordered.  We believe this explains previous numerical
work on this problem which reported the persistence of Ising-like
order at small $\sigma$.  In the most recent of these works, 
Cataudella \cite{R5}
reports at $\sigma\simeq 0.113$ a finite $T_c$ to an
Ising-like ordered state.  The correlation length exponent that he
finds is $\nu\sim 1.7$, clearly
different from that of the pure model.  Using our scaling form (iii)
we can estimate that at this value of $\sigma$, $\xi\sim 120$,
much larger than Cataudella's largest system size of $L=36$.
His results may thus be reflecting a cross over
region at $L<\xi$, rather than a true transition.

We thank Prof. Y. Shapir for many valuable discussions.
This work has been supported by DOE grant DE-FG02-89ER14017.


\begin{thebibliography}{99}

    \bibitem[\dag]{R0}Present address: Department of Physics and 
         Astronomy, McMaster University,  Hamilton, Ontario, L8S 4M1 
         Canada

	\bibitem{R1}  E.~Granato and J.~M.~Kosterlitz, Phys. Rev. B {\bf 
            33}, 6533 (1986); M.~Rubinstein, B.~Shraiman and 
            D.~R.~Nelson, Phys. Rev. B {\bf 27}, 1800 (1983).

	\bibitem{R2}  T.~Nattermann, S.~Scheidl, S.~E.~Korshunov and M.~S.~Li, 
             J. Phys. (France) I {\bf 5}, 565 (1995); L.-H.~Tang, Phys. 
             Rev. B {\bf 54}, 3350 (1996); S.~Scheidl, Phys. Rev. B {\bf 55}, 
             457 (1997); M.~C.~Cha and H.~Fertig, Phys. Rev. Lett. {\bf 74}, 
             4867 (1995); J.~Maucourt and D.~R.~Grempel, Phys. Rev. 
             B {\bf 56}, 2572 (1997); Y.~Ozeki and H.~Nishimori, J. 
             Phys. A {\bf 26}, 3399 (1993).

	\bibitem{R6}  M.~G.~Forrester, S.~P.~Benz and C.~J.~Lobb, Phys. Rev. 
             B {\bf 41}, 8749 (1990); A.~Chakrabarti and C.~Dasgupta,  
             Phys. Rev. B {\bf 37}, 7557 (1988); S.~P.~Benz, M.~G.~Forrester, 
             M.~Tinkham and C.~J.~Lobb, Phys. Rev. B {\bf 38}, 2869 (1988);
             M.~G.~Forrester, H.~J.~Lee, M.~Tinkham 
             and C.~J.~Lobb, Phys. Rev. B {\bf 37}, 5966 (1988); S.~E.~Korshunov, 
             Phys. Rev. B {\bf 48}, 1124 (1993).

	\bibitem{R20}  J.~M.~Kosterlitz and M.~Simkin, Phys. Rev. Lett. {\bf 
             79}, 1098 (1997).

	\bibitem{R7}  J.~Villain, J. Phys. C. {\bf 10}, 1717 and 4793 (1977).

	\bibitem{R3}  S.~Teitel and C.~Jayaprakash, Phys. Rev. B {\bf 27}, 
            598 (1983); Phys. Rev. Lett. {\bf 51}, 1999 (1983).

	\bibitem{R4}  E.~Granato and J.~M.~Kosterlitz, Phys. Rev. Lett. {\bf 
            62}, 823, (1989); M.~Y.~Choi, J.~S.~Chung and D.~Stroud, Phys. 
            Rev. B {\bf 35}, 1669 (1987).

	\bibitem{R5}  V.~Cataudella, Europhys. Lett. {\bf 44}, 478 (1998).

    \bibitem{R7.5} J.~Villain, J. Physique {\bf 36}, 581 (1975).

	\bibitem{R8}  E.~Fradkin, B.~Huberman and S.~H.~Shenker, Phys. Rev. B 
            {\bf 18}, 4789 (1978); A. Vallat and H. Beck, Phys. Rev. B 
            {\bf 50}, 4015 (1994).

	\bibitem{R18}  P.~Olsson, Phys. Rev. B {\bf 52}, 4511 (1995).

	\bibitem{R9}  Y.~Imry and S.-K.~Ma, Phys. Rev. Lett. {\bf 35}, 1399 
             (1975).

	\bibitem{R10} C.~Denniston and C.~Tang, Phys. Rev. Lett. {\bf 79}, 
             451 (1997) and Phys. Rev. B {\bf 58}, 6591 (1998).

	\bibitem{R11}  T.~C.~Halsey, J. Phys. C {\bf 18}, 2437 (1985).

	\bibitem{R12}  J.-R. Lee and S. Teitel, Phys. Rev. B {\bf 46}, 3247 (1992).

	\bibitem{R16}  K.~Binder, Z. Phys. B {\bf 50}, 343 (1983).

	\bibitem{R13}  M. Aizenman and J. Wehr, Phys. Rev. Lett. {\bf 62}, 
            2503 (1989); K. Hui and N. Berker, {\it ibid.} 
            2507.

	\bibitem{R15}  P.~Olsson, Phys. Rev. B. {\bf 55}, 3585 (1997).

	\bibitem{R17}  R.~N.~Bhatt and A.~P.~Young, Phys. Rev. B {\bf 37}, 
            5606 (1988).

	\bibitem{R19}  P.~Gupta, S.~Teitel and M.~J.~P.~Gingras, Phys. Rev. 
            Lett. {\bf 80}, 105 (1998).

\end{thebibliography}
\end{document}